\documentclass[acmsmall]{acmart}

\setcopyright{rightsretained}
\acmJournal{PACMHCI}
\acmYear{2024} \acmVolume{8} \acmNumber{CSCW2} \acmArticle{446} \acmMonth{11}\acmDOI{10.1145/3686985}

\usepackage{hyperref}
\usepackage[noabbrev,capitalise,nameinlink]{cleveref}

\hypersetup{colorlinks={true},linkcolor={blue}}
\usepackage{tabularx}
\usepackage{xcolor, colortbl,booktabs}  
\usepackage[first=0,last=9]{lcg}

\begin{document}

\title[Citizen-Led Personalization of User Interfaces]{Citizen-Led Personalization of User Interfaces: Investigating How People Customize Interfaces for Themselves and Others}


\author{Sérgio Alves}
\email{sfalves@fc.ul.pt}
\orcid{0000-0002-3647-0624}
\affiliation{%
  \institution{LASIGE, Faculdade de Ciências, Universidade de Lisboa}
  \city{Lisbon}
  \country{Portugal}
}

\author{Ricardo Costa}
\email{fc55106@alunos.fc.ul.pt}
\orcid{0000-0001-7234-7984}
\affiliation{%
  \institution{LASIGE, Faculdade de Ciências, Universidade de Lisboa}
  \city{Lisbon}
  \country{Portugal}
}

\author{Kyle Montague}
\email{kyle.montague@northumbria.ac.uk}
\orcid{0000-0002-9589-9471}
\affiliation{%
    \institution{Computer and Information Sciences, Northumbria University}
    \city{Newcastle upon Tyne}
    \country{United Kingdom}
}

\author{Tiago Guerreiro}
\email{tjvg@di.fc.ul.pt}
\orcid{0000-0002-0333-5792}
\affiliation{%
  \institution{LASIGE, Faculdade de Ciências, Universidade de Lisboa}
  \city{Lisbon}
  \country{Portugal}
}

\renewcommand{\shortauthors}{Alves, S. et al.}

\begin{abstract}
  User interface (UI) personalization can improve usability and user experience. However, current systems offer limited opportunities for customization, and third-party solutions often require significant effort and technical skills beyond the reach of most users, impeding the future adoption of interface personalization. In our research, we explore the concept of UI customization for the self and others. We performed a two-week study where nine participants used a custom-designed tool that allows websites' UI customization for oneself and to create and reply to customization assistance requests from others. Results suggest that people enjoy customizing for others more than for themselves. They see requests as challenges to solve and are motivated by the positive feeling of helping others. To customize for themselves, people need help with the creative process. We discuss challenges and opportunities for future research seeking to democratize access to personalized UIs, particularly through community-based approaches.
\end{abstract}



\received{July 2023}
\received[revised]{January 2024}
\received[accepted]{March 2024}

\maketitle

\section{Introduction}
Traditionally, designers fully control the user interface (UI) crafting process. They devise designs, define components, run experiments, or collect and assess interaction data at will, intending to create one-size-fits-all solutions aimed at an average user \cite{saati2005towards}. However, one-size-ﬁts-all UIs cannot handle the context variability that leads to an unpleasant user experience \cite{hussain2018model}, and, today, UIs can still be a factor of exclusion \cite{msweli2020enablers}. Ultimately, each user possesses unique characteristics that affect their experience and performance when interacting with UIs (e.g., technological proficiency, self-confidence) \cite{Sundar2010}; but limited options to completely decide or adjust how interactive systems are presented.

UI customization empowers individuals with the possibility to adjust UIs to their needs, fixing or improving (in someone's eyes) the original design. Customization involves manipulating properties and values, such as color, size, or position, of visual elements, including buttons, headers, paragraphs, or images. The result can be a visually adapted menu, data feed, or a new color scheme (e.g., \cite{nebeling2013crowdadapt, kumar2011bricolage}).

People benefit from customization regardless of skills, age, or expertise. Customized UIs can improve user experience \citep{nebeling2013crowdadapt}, efficiency \citep{reinecke2011improving, reinecke2011improving}, overall satisfaction \citep{10.1145/3457151, gajos2010automatically, 10.1080/07421222.2015.1029394}, website stickiness \citep{10.1080/07421222.2015.1029394}, or enable access to initially inaccessible areas of the screen \citep{wu2022reflow}. Simultaneously, customization yields psychological benefits by enhancing the senses of control \citep{marathe2011drives}, identity, or personal agency \citep{sundar2008self}. Overall, although people are interested in making devices look and feel ``their own'' \citep{hakkila2006personal}, their primary motivation to customize lies in the pursuit of efficiency \citep{mackay1991triggers}.

Previous work has limited presence in current UIs. When users open a web page or application, they often encounter minimal options to adjust the UI to their abilities and likings (e.g., changing color schemes or font size). Existing customization solutions \cite{tampermonkey, stylish} require technical skills, time, and effort that not everyone can provide \cite{mackay1991triggers}. Moreover, complex problems can require more knowledge than a single person possesses \citep{paterno2013end}. Consequently, tech-savvy users feel comfortable customizing, while the less tech-savvy have negative attitudes towards customizable interfaces \cite{Sundar2010}.


The ultimate goal of this work is to democratize access to customization and customized UIs, providing customization benefits to ordinary citizens who may lack the availability or expertise to customize.

We have witnessed the democratization process of interactive technologies being made through the power of the open community, for instance, with Social Accessibility \cite{takagi2008social} or Thingiverse \cite{buehler2015sharing} in the field of accessible technologies. There have also been efforts to give agency to individuals by allowing them to develop \cite{lieberman2006end} or commission their own interactive systems \cite{garbett2016app,vlachokyriakos2016digital}. This democratization process enables the rapid development of low-cost technology \cite{buehler2015sharing} while improving technological support for individual citizen needs and their sense of agency over those technologies. In the UI customization field, community-based approaches that foster relations between individuals with different roles, skills, or expertise are under-explored. A gap that persists despite recent evidence suggesting that non-expert users expect assistance in implementing customization details \citep{10.1145/3491102.3501931}.


In this work, we explore the concept of end-user customization of UIs for the self and, for the first time, for others. We introduced the concept of requesting customization assistance to enable users to customize for each other, exploring mutual help mechanisms between users. This work is a first step in introducing community-based components in UI customization. We defined the following research questions: \textit{(1) How do people independently use a customization tool to customize for the self -- including context, challenges, and motivations (\textbf{RQ1})? (2) How do they make use of the chance to request customization assistance (\textbf{RQ2})? (3) How do they react to possibly supporting others to customize (\textbf{RQ3})?}

We developed a browser extension, GitUI, that enables users to customize existing web pages and apply those changes to the current and future usage sessions. Users can request customization assistance when needed and assist others. We present an exploratory study where nine participants, experts and non-experts, used GitUI for two weeks, customizing at will and making and replying to customization assistance requests from the crowd.

Results suggest that, regardless of expertise, people enjoy customizing for others more than for themselves. To customize for themselves, people need help with the creative process, including guidance to understand what customization operations they can perform and what would benefit them (\textbf{RQ1}). To customize for others, people only need to focus on developing the solution. They were motivated by the challenge it represents (participants saw an assistance request as a challenge to solve) and the positive feeling of helping others (\textbf{RQ3}). Future solutions should leverage these motivation factors to ensure long-term user engagement by increasing recognition, feedback, and user communication, as well as introducing gamification mechanisms. We also identified that factors like \textit{self-efficacy}, \textit{sense of control}, or \textit{selflessness} are challenges to requesting customization assistance (\textbf{RQ2}). Overall, participants assumed to be in favor and motivated to be part of a community that personalizes.

We contribute with (1) an in-depth analysis of how people customize for themselves and others; (2) the design and users' feedback of a working prototype demonstrating the feasibility of end-user UI customization for the self and others; and (3) a discussion that should inform developers of future community-based UI personalization solutions.

\section{Background}
We analyzed previous work on UI personalization and community-based solutions to improve UIs.
 
\subsection{Personalization of User Interfaces}
We first delve into the diverse dimensions of tailoring UIs, examining the personalization spectrum and the factors impacting personalization.

\subsubsection{The Personalization Spectrum}
Personalization is divided into adaptation, driven by the system \cite{gajos2017influence, kumar2011bricolage}, and customization, done by the user \cite{nebeling2013crowdadapt, bila2007pagetailor, hakkila2006personal}. These forms of personalization represent external run-time repair solutions to adjust UIs, differing from adaptive \cite{kuhme1993user, brusilovski2007adaptive}, adaptable \cite{10.1145/1166253.1166301}, or automatically generated UIs \cite{gajos2010automatically}.

Adaptation relies on predefined rules and user data to automatically adjust UIs. For instance, Split Adaptive Interfaces \citep{gajos2017influence} predict relevant functionalities, moving them to an adaptive shortcut toolbar. On the other hand, customization allows users to modify UIs directly. A good example is CrowdAdapt \cite{nebeling2013crowdadapt}, a web direct manipulation toolkit that supports seven operations: move, resize, spacer, hide, collapse, font size, and multi-column. It also contains a crowd-based component that allows users to use crowd adaptations; however, they can not interact with each other. A study with 93 participants revealed the potential for enhancing the browsing experience through customization. Both personalization approaches have pros and cons. Customization does not need to collect personal data and generally allows more fine-tuned personalizations. However, it requires users to invest significant time, which may outweigh customization benefits \cite{bunt2007supporting}. People tend to customize only when it is worth the effort \cite{mackay1991triggers}. Adaptation reduces the workload, although it does not leave the user in complete control and raises privacy concerns \cite{Sundar2010}.

Several strategies attempt to reduce the customization effort while providing more control than adaptation solutions. Mixed-initiative approaches, which still do not allow users complete control, combine both advantages by proposing suggestions of personalizations that users can accept/reject \cite{10.1145/3457151, bunt2007supporting}-- based on rules and expected usage. Alternatively, example-based approaches like Bricolage \cite{kumar2011bricolage} allow the transfer of design and content between web pages by creating coherent mappings between similar elements. Human-generated mappings, collected using crowd-sourcing, automatically transfer the content from one page into the style and layout of another. UI retargeting \citep{10.1145/3411764.3445573, 10.1145/3472749.3474796} systems follow a similar approach -- although they mainly were studied to support the work of designers. For instance, Umitation \citep{10.1145/3472749.3474796} supports extracting and retargeting dynamic UI behavior examples from existing to new (target) websites. While ``borrowing'' UI behaviors, users can also adjust UI design properties such as width or opacity.

More recently, \citet{10.1145/3491102.3501931} presented Stylette, which allows website customization using natural language. Users can use their voice to express their goals and interact with a set of design alternatives presented by the system. In a first study with eight non-experts, where participants could verbally request to a researcher styling changes, the authors found that novice users provided customization requests frequently vague (lacking specific details or using abstract terms) and that they did not want to dedicate the mental effort reasoning about the details -- expecting the support from an expert instead. Results from a second study suggest that voice interaction started to limit participants' productivity as they acquired more knowledge with Stylette (people gradually sought more direct interaction methods).

\subsubsection{Factors Impacting Personalization}
Multiple human factors influence the customization process at different stages. Exposure and awareness of customization features and social influence trigger a desire to customize \citep{banovic2012triggering}. Additionally, there is a relationship between customization and the senses of control and identity \citep{marathe2011drives}, as well as the sense of personal agency, by making UIs present more relevant content \citep{sundar2008self}. While customizing, locus of control (the extent to which people believe they can control events affecting them) also affects customization effectiveness \citep{lalle2019role}.

Individuals' response to customization depends on their tech-savviness. Less tech-savvy users have negative attitudes toward a UI when asked to customize it but a positive attitude when accessing an already personalized UI \citep{Sundar2010}. On the other hand, compared to the less tech-savvy, tech-savvy users showed more positive attitudes and a higher sense of control when allowed to customize.

The 5-factor personality traits (openness, conscientiousness, extroversion, agreeableness, and neuroticism) influence how people customize or use personalized UIs \citep{saati2005towards}. For instance, there is a correlation between color preferences and extroversion, conscientiousness, and openness. Other personality traits affecting personalization are the need for cognition, uniqueness, and variety-seeking \citep{10.1145/1453794.1453800} -- people with a higher need for cognition or uniqueness value personalization more than variety-seekers.

In summary, these works suggest that people are interested in and benefit from personalization; however, today's UIs still have little support for more tuned personalization (e.g., move elements). Customization is allowed by operating systems (e.g., iOS \cite{appleSupport} individual application preferences), browsers (e.g., zoom in/out), and applications (e.g., color modes). Still, existing built-in customization operations are restrictive, and third-party solutions have failed to become an active part of digital life. Our work focuses on improving and democratizing the UI customization process for ordinary citizens while reducing the effort barrier\cite{mackay1991triggers}.

\subsection{Community-based Solutions to Improve User Experience}
Although one goal of personalization is inclusiveness, not everyone is skillful enough to customize or wants to invest time doing it \cite{mackay1991triggers}. Community-based solutions, common in other research fields, can be a solution to deliver personalization to these people; however, they still need to be explored.

CrowdUI \cite{10.1145/3394978} allows website community members to visually express their design improvement needs and inform the website's owner. A study showed that beginner users were more open than experts to using the tool. \citet{10.1145/2494603.2480319} presented a mechanism for crowdsourcing UI adaptations. Enterprise users can adapt a UI, and an administrator checks, integrates, and publishes the adapted UI. However, these are not citizen-led approaches, limiting end-users' control over the process and design decisions.

Another critical concept is crowdsourced Web site components \cite{nebeling2012crowdsourced}. The idea is to build components that continuously improve its content, presentation, and behavior with the help of the crowd. Non-experts can build their components on top of others created by more experienced users. This concept led to CrowdAdapt \cite{nebeling2013crowdadapt}. CrowdAdapt is, to the best of our knowledge, the only customization tool (that requires no scripting) that allows users to obtain adaptations made by others. The system automatically shares created adaptations with other users and applies them according to their settings when visiting a web page. Users can also preview different adaptations. Overall, users enjoyed the concept.

Two mainstream solutions, Stylish \cite{stylish} and Tampermonkey \cite{tampermonkey}, allow users to create and share CSS and JavaScript adaptations, respectively. However, in both, users must write code to customize without customization assistance, yet it is possible to install adaptations of others. A successful example of crowd-based UI adaptations comes from the gaming community, where players customize game UIs and share them with the rest of the community \cite{dyck2003learning}.

In these works, there is no communication or mutual help between users. Users are either not in control of the customization process (for instance, when installing adaptations from the crowd), or, if they want that control, they may not possess the required skills or availability to customize. Previous work \cite{10.1145/1753326.1753357, 10.1145/2506364.2506368} showed that non-experts of the crowd can provide proper aesthetic preference judgments. We expand previous work by studying how human-powered mechanisms can provide direct customization assistance to those lacking the skills or time to do so. By requesting assistance, users can access a customized UI, decide, and have control over that customization, but provide less effort.

We have seen the successful application of these concepts in accessible computing. \citet{takagi2008social} introduced the concept of Social Accessibility. The idea is to make existing content accessible by using the power of the open community. When users encounter an accessibility problem, they can report it to a social computing service. Volunteers then discuss, create, and publish a fix. Other examples include the Social CheatSheet \cite{10.1145/3134737}, a UI overlay with step-by-step tutorials curated by the community, and RISA \citep{rodrigues2021promoting}, a human-powered task assistant for smartphone users.

With our work, we aim to explore how people customize for themselves and, mainly, to understand how they react to the introduction of crowd-based components, where, for the first time (to the best of our knowledge), they can request customization assistance from others and help others to customize. Compared with existing approaches, our solution provides benefits equivalent to existing customization solutions while mitigating its primary limitations. We enable users to completely tailor UIs to their needs (contrary to automated solutions), contributing to the senses of control, identity, or personal agency while reducing the required effort, time, and skills \cite{mackay1991triggers} -- which results in negative attitudes from less tech-savvy users\citep{Sundar2010} and reliance on experts for assistance \citep{10.1145/3491102.3501931}.  In contrast to existing approaches, tech-savvy users can also benefit from the community by seeking assistance for complex issues beyond an individual's expertise \citep{paterno2013end}. Our approach depends on people being available to communicate with and help others. Previous work shows that people are available to contribute with improvement ideas \cite{10.1145/3394978}, generate adaptations for the crowd \cite{nebeling2013crowdadapt}, or communicate to increase UIs accessibility \citep{takagi2008social}. Our ultimate goal is to identify new opportunities to democratize, through community-based approaches, access to customized UIs for those lacking availability or skills.

\section{Methods}
\begin{figure*}
	\centering
	\includegraphics[width=1\textwidth]{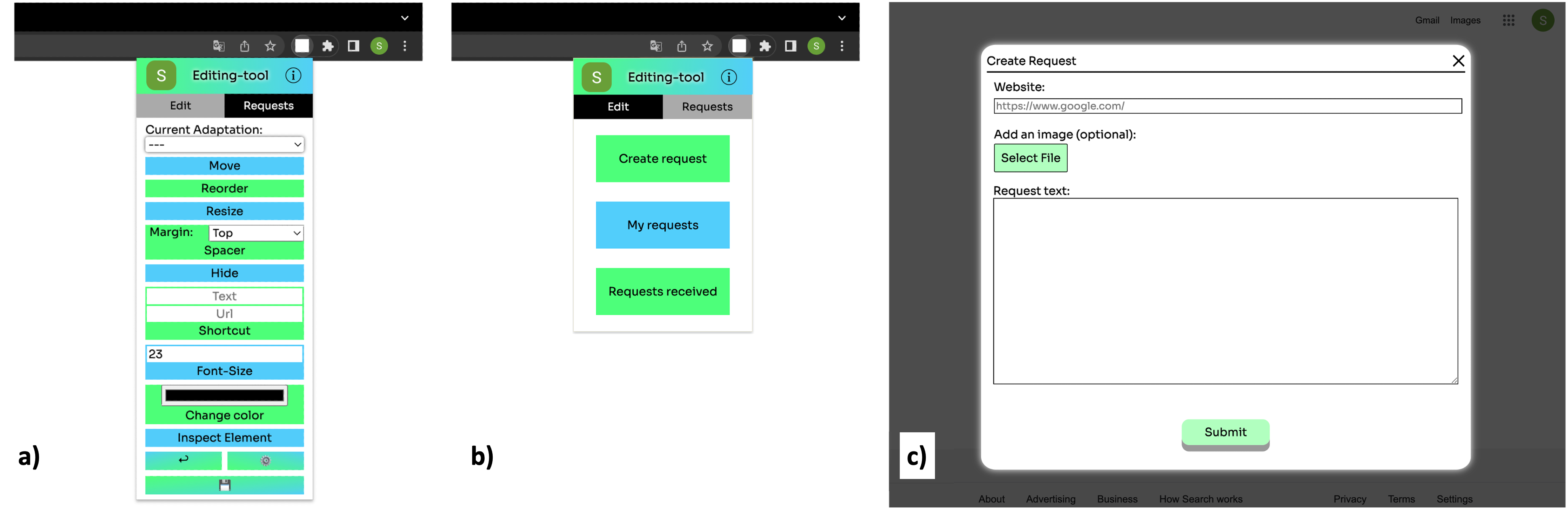}
	\caption{The GitUI extension allows users to customize with nine operations (a), and manage (b) and create (c) customization requests.}
	\label{fig:tool}
	\Description{Three images with Google Chrome Browser with the google.com page open. On the left image (a): on the top bar of the browser, the extension popup is open. From the top to the bottom: A title saying editing tool, and next to it an icon for more information; two tabs (Edit and Requests); a dropdown saying current adaptation; a button with the text move; a button with the text reorder; a button with the text resize; a button with the text spacer (and a dropdown to select the margin); a button with the text Hide; two inputs (one for text, other for URL) followed by a button with the text shortcut; an input with a value of 23 followed by a button with the text Font-size; a colour input (black) followed by a button with the text Change Color; a button with the text Inspect element; two buttons (one back arrow on the left, and a settings icon on the right); and a save icon.}
\end{figure*}

We performed a two-week exploratory study where participants used a custom-designed tool, GitUI (\cref{fig:tool}). We studied the concept of end-user customization of UIs for the self and others. Our goals were to understand (1) how people independently use a customization tool -- including context, challenges, and motivations; (2) how they benefit from requesting customization assistance; and (3) how they react to the possibility of assisting others. Overall, we wanted to inform and explore opportunities for future community-based personalization approaches, detecting challenges, motivations, and possible benefits. The study followed all ethical considerations required by our university.

\subsection{Procedure}
We divided the study into three steps: (1) a session to familiarize participants with the study and GitUI; (2) a two-week period where they used GitUI; and (3) an interview session. We collected feedback on how to improve GitUI; however, our main goal was not to study its usability. Instead, we wanted to understand how people react to the concept of requests (creating and replying) and how that interconnects with the customization for the self, including contexts of use, motivations, or challenges. Sessions were audio recorded and held in a private room inside our university.

\subsubsection{Think-Aloud Training Session}
In this phase ($\approx$1h), a researcher trained participants in using GitUI. We wanted to ensure that they understood the concept of customization, GitUI's operations, and the requests workflow.

Participants had to customize a website to look as close as possible to a provided final design (exploring all the customization operations available). This final design included a before-after image circling around components to change. We started by showing participants how to access GitUI in the browser and allowing them to complete the task without any specific order. We asked them to express their thoughts and perspectives about the system. The research team prepared an interview script containing elicitation questions to extract their perspectives when necessary. Participants should explore the tool independently, recurring to researchers' assistance when necessary. They used the research team's laptop to complete the task. This procedure lasted between 11.8 and 32.6 minutes (21.5±8 minutes).

Next, participants explored the requests. We clarified that they could create requests when they lacked the desire, availability, or expertise to customize themselves. We verbally gave participants examples of what and how they could request and guided them through GitUI to demonstrate the process. We also demonstrated how to reply to requests. The research team did not set any guidelines regarding how to make or reply to requests. We also informed participants that their requests would be anonymous to other users and not to be concerned with request complexity.

After the task, participants completed the NEO-FFI-20 (20 Likert scale questions (0–4)) \cite{bertoquini2006} and a computer self-efficacy scale (10 Likert scale questions (1–10)) \cite{10.2307/249688}. The NEO-FFI-20 is a self-reported questionnaire that assesses the big five personality traits, which influence the customization for the self \cite{saati2005towards}. We wanted to study if these traits or the computer self-efficacy of participants also impact how they customize or reply to customization requests.

\subsubsection{Customization at Home}
Participants should use GitUI for two weeks on at least three websites per week, either by customizing for themselves or creating an assistance request. Replying to requests does not count towards this minimum goal. We did not provide instructions regarding what and how to personalize, nor did we force participants to decide between customizing or asking for assistance. We pretended to create an experience as close as possible to an authentic one. We asked participants to act naturally and not force themselves to customize or create requests. We kept logs of their interactions with GitUI, including the frequency of use, customized websites, operations performed, or created and replied assistance requests. By providing them with equal opportunity to request assistance or customize themselves, we also wanted to understand the context and motivations for whether they prefer one or another.


The requests workflow was the following. When someone made a request, the research team manually assigned it to another participant (to ensure proper workload distribution). Each participant received a new request every two or three days. Participants could respond to these within 48 hours. If requesters did not obtain a reply, the research team ensured one. If there were no requests after the second day, and to ensure the flow of the study, participants received fictional requests. Every time there was a new request, participants were notified by email. Together with the request, they also received fictional information about the requester (name, age, and profession) to understand if it influences the replies. Participants did not receive any instructions for executing the requests. When participants did not reply to a request, they received an email informing them that the system assigned the request to another user. We only informed participants of fictional requests at the end of the study.

Fictional requests were rationally created by the research team before the study, exploring different personas, categories of websites, and complexity. In total, we pre-made 12 fictional requests. We internally labeled four requests with a difficulty of one, five with a difficulty of two, and three with three. The difficulty depends on the number and complexity of the necessary customization operations. Based on personas' age and profession, we had three requests that participants could interpret as deriving from tech-savvy people and four from non-tech-savvy. Three websites were from social networks and entertainment, five from information, and four from health or governmental resources.

Requests had different motivations, which we used to encourage people to explore different operations (e.g., color changes for accessibility or shortcut creations for efficiency). Participants received at least one request from each difficulty, website category, and tech-savvy level. We used a spreadsheet to record requests assignment, including dates or status, and manage subsequent assignments (based on the categories, date, and number of assignments).

An example of a simple customization request on a social media website\footnote{https://twitter.com} is: ``\textit{I do not care about the menu on the left. I just need a button to refresh the tweets on the page (Sarah, 60 years old, Lawyer)}.''

More advanced requests could require code writing: ``\textit{I enjoy Reddit's \footnote{https://www.reddit.com} interface; however, one thing bothers me: the comments on a post are expanded by default when you open it. I usually only want to read the main comments, not the replies. I prefer having the replies hidden by default with an option to visualize them (John, 35 years old, Designer)}.''

In total, we assigned 38 requests, 36 of them fictional. Participants replied to 31 (82\%) of the 38 requests made.

\subsubsection{Final Interview}
We concluded the study with an individual interview about participants' general experience with GitUI, context of use, motivation, goals, and future personalization opportunities.

\subsection{Sample and Recruitment}
We recruited people with different technical and demographic profiles. In particular, we aimed to recruit expert and non-expert users, to understand their responses to the roles of requester and volunteer. We recruited participants using the university mailing lists and social networks. Participants should be internet users for more than four hours a week. They indicated interest by completing an online questionnaire. We rewarded participants with a voucher of €20.

In total, nine participants (P1-P9), aged between 25 and 59 years (31.8±10.4 years), concluded the study (\cref{table:participants}). Two weeks before the main study, a pilot study with P1 and P2 ensured that our instrument worked and helped improve GitUI's usability and stability. This procedure was similar to the main study. First, P1 and P2 performed the think-aloud training, which allowed us to estimate the session duration better, test GitUI, and adjust our interview script. Then, we instructed P1 and P2 to use GitUI at home for one week but without minimum requirements. They were aware of the pilot study but not the number of participants. Finally, we interviewed P1 and P2. We decided to include P1 and P2 in the results of our study as they still used GitUI enough times to provide valuable input.

\begin{table}[h]
\centering
\caption{Participants profile, where HBI (hours a day browsing on the internet), CSE (computer self-efficacy), \#Customization (total customization templates created for the self), \#Requests (total requests), \#Replies (total replies submitted for total requests received).}
\label{table:participants}
\Description{
For P1: Age 25, Expert Yes, BI 8, CSE 8, #Customization 1, #Requests 0, #Replies 4/4.
For P2: Age 33, Expert Yes, BI 12, CSE 10, #Customization 0, #Requests 0, #Replies 4/5.
For P3: Age 35, Expert No, BI 9, CSE 8.3, #Customization 2, #Requests 2, #Replies 4/5.
For P4: Age 59, Expert No, BI 9, CSE 5.6, #Customization 7, #Requests 0, #Replies 4/4.
For P5: Age 31, Expert Yes, BI 12, CSE 6.7, #Customization 8, #Requests 0, #Replies 1/4.
For P6: Age 28, Expert Yes, BI 10, CSE 10, #Customization 6, #Requests 0, #Replies 3/3.
For P7: Age 28, Expert Yes, BI 12, CSE 6.6, #Customization 6, #Requests 0, #Replies 3/4.
For P8: Age 25, Expert Yes, BI 9, CSE 8.5, #Customization 6, #Requests 0, #Replies 3/4.
For P9: Age 23, Expert No, BI 1, CSE 7, #Customization 6, #Requests 0, #Replies 5/5.}
\begin{tabular}{c|c|c|c|c|c|c|c}
\toprule
\textbf{ID} & \textbf{Age} & \textbf{Expert} & \textbf{HBI} & \textbf{CSE} & \textbf{\#Customization} & \textbf{\#Requests} & \textbf{\#Replies}\\
\midrule
\textbf{P1} & 25 & \checkmark & 8 & 8 & 1 & 0 & 4/4\\
\rowcolor{lightgray!50}
\textbf{P2} & 33 & \checkmark & 12 & 10 & 0 & 0 & 4/5\\
\textbf{P3} & 35 & - & 9 & 8.3 & 2 & 2 & 4/5\\
\rowcolor{lightgray!50}
\textbf{P4} & 59 & - & 9 & 5.6 & 7 & 0 & 4/4\\
\textbf{P5} & 31 & \checkmark & 12 & 6.7 & 8 & 0 & 1/4\\
\rowcolor{lightgray!50}
\textbf{P6} & 28 & \checkmark & 10 & 10 & 6 & 0 & 3/3\\
\textbf{P7} & 28 & \checkmark & 12 & 6.6 & 6 & 0 & 3/4\\
\rowcolor{lightgray!50}
\textbf{P8} & 25 & \checkmark & 9 & 8.5 & 6 & 0 & 3/4\\
\textbf{P9} & 23 & - & 1 & 7 & 6 & 0 & 5/5\\
\rowcolor{lightgray!50}
\bottomrule
\end{tabular}
\end{table}

\subsection{Apparatus}
We developed a Google Chrome extension (\cref{fig:tool}) that participants could freely use to create and apply customization templates on any website and to create and reply to customization requests. A template results from a set of customization operations, which can be updated or submitted as a reply. Templates can be specific to a web page or the whole website domain. Once saved, users can activate the template in the tool's menu. When activated, GitUI applies the template whenever the user visits the target web page.

Users can act as requesters or volunteers, depending on whether they are making or replying to a request. An assistance request consists of a website URL and a description, which, in the study context, the research team manually assigns to a volunteer. The volunteer then replies by submitting a customization template for the requester's use.

Users interact with GitUI through a popup menu that opens when they click the extension icon. The menu is divided into the \textit{Customization} and the \textit{Request} tabs. We developed GitUI using JavaScript, HTML, and CSS. It utilizes Firebase as the storage platform, and users log in with a single click using the account associated with their browser.

\subsubsection{Customization}
The \textit{Customization} tab (\cref{fig:tool} a)) allows to create and apply customization templates. The customization workflow is as follows: (1) if necessary, users start by defining the values (e.g., color) on the popup menu of the customization operation to apply; (2) they click on the operation name; and (3) they use the mouse to select a web page element to apply the operation (the system highlights elements on hover). To finish, they save the template and define a title and a description. Templates are associated with one user and can be activated (using the \textit{dropdown} on top) and updated at any time. If none is active, users interact with the original UI. They can access a template list when visiting the target website or replying to a request.

GitUI supports nine operations (motivated by previous work \cite{nebeling2013crowdadapt, nebeling2011tools}) and JS or CSS injection into any web page. Users can apply operations with a single click, like the \textit{Hide}, \textit{Font-Size}, \textit{Change Color}, or \textit{Shortcut} (which, given a text and URL, inserts a link) or perform a 3-step process (where they adjust the element using the cursor and then press \textit{Enter} to confirm) for the \textit{Move}, \textit{Reorder}, \textit{Resize}, and \textit{Spacer}. They can revert operations or press \textit{Esc} to interrupt them. The \textit{Inspect Element} and the \textit{Advanced Editing} modal (with CSS and JS syntax checkers) complement these operations.



\subsubsection{Requests}
The \textit{Request} tab (\cref{fig:tool} b)) enables users to create customization requests and view a list of the created requests and another with the received ones. There is no direct communication between users: (1) a requester submits a request; (2) the research team assigns it to a volunteer; (3) the volunteer creates a customization template for an anonymous requester; (4) the requester can access and use it.

To create an assistance request, users fill out a form (\cref{fig:tool} c)). They indicate the target website, write a detailed explanation of their wish, and, optionally, provide an image with visual cues. Users can verify the status of their created requests in the ``My Requests'' list. When a volunteer solves a request, GitUI notifies users through email and a notification icon. Requesters then access the ``My Requests'' list to download the reply template. From there, requesters can use the reply template as if they created it (e.g., further customize it).

When users receive an assistance request, we notify them through GitUI and email. To reply to a request, users create a customization template (as usual) and submit it as a reply. To submit a reply, users open the received requests list, click on the request, and, in a \textit{dropdown}, select one of the customization templates available for that website. In the study context, the research team verifies the reply and, if acceptable, makes it available to the requester.

\subsection{Analysis}
The two researchers who conducted the sessions analyzed the interviews. We performed a thematic analysis \cite{braun2006using} following the first three stages outlined by \citet{HALCOMB200638}, which do not require transcription of interview recordings. First, during all interviews, while one researcher conducted the interview, the other took detailed notes of relevant phrases, recurring ideas, and non-verbal cues, such as frustration and excitement. After each interview, the researchers held debriefing sessions. After the study, both researchers reviewed the audio records in consultation with the notes. They amended the notes to ensure that they accurately reflected the data. Relevant quotes to interpretation were carefully transcribed and examined. Then, we individually identified key codes across interviews and grouped codes representing similar phenomena into themes. In two group sessions, both researchers discussed, merged, and reviewed the themes to ensure they captured the depth and range of data collected across interviews.

For quantitative data, we calculated the mean, standard deviation, minimum, and maximum for continuous variables; and created tables of frequencies for nominal variables. We calculated the the \textit{Pearson product-moment correlation coefficient} between each personality trait, collected through the NEO-FFI-20 questionnaire (\cref{ipipneo}), and the participants' activity during the study (\cref{table:participants}), but found no relationship between them.

\section{Results}
In this section, we present study results grouped by the three components we investigate: customization for the self, assistance requests, and customization for others. We conclude by reporting people's feedback on GitUI.

\subsection{Customizing for the Self}
Customization for the self was generally limited to the minimum (\cref{table:participants}). Next, we describe the participants' experience.

\subsubsection{Motivation, Context, and Benefits}
Participants aimed to \textbf{simplify} and \textbf{optimize} web pages: \textit{``my goal was to eliminate the noise... to make websites simpler and more objective... add shortcuts''} (P4). Customization decisions were informed and made on websites that people visit more often: \textit{``I know what I need in the websites I visit most (...) keep and reorder the most relevant information (...) I think `this is what I want and I do not need distractions'\,''} (P9).

Created templates did not have an aesthetic focus except for creating dark themes. P8 commentated: \textit{``I would only adapt aesthetics to be more efficient... which is my focus... but the design of the websites I visit is good enough''}.

\begin{figure*}
	\centering
	\includegraphics[width=1\textwidth]{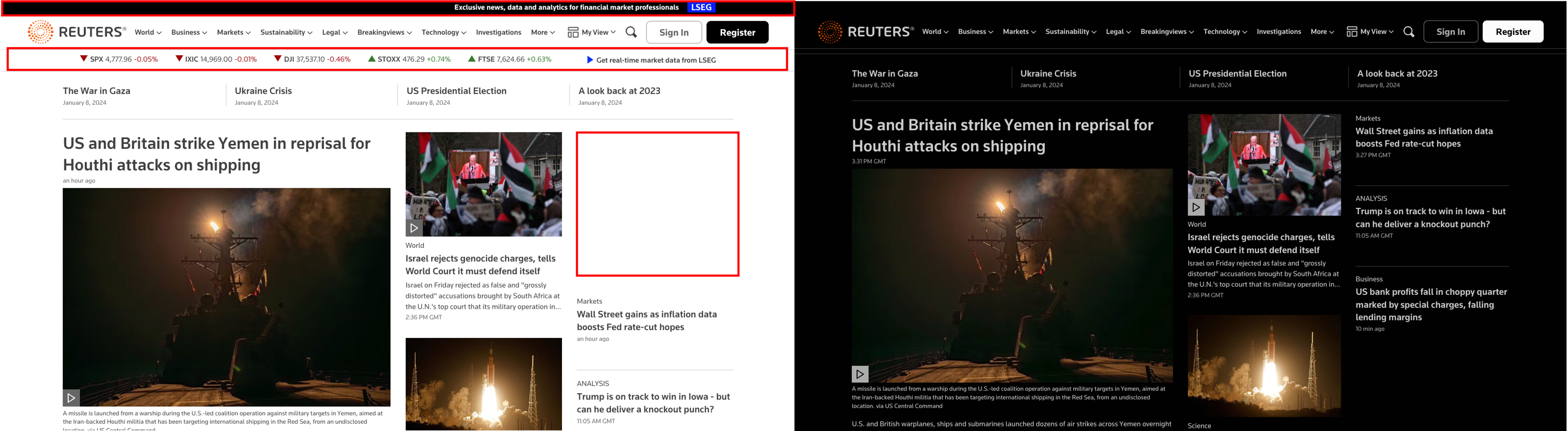}
	\caption{Example of a customization template: the original (left) and the customized version (right). P6 customized the colors, margins, and hid elements (hidden elements highlighted in red).}
	\label{fig:example}
	\Description{Two screenshots of the same news website. One the left with a light background, dark font, icons on the menu, and advertisements. On the right, with dark background, light fonts, no icons on the menu and no advertisements.}
\end{figure*}

Four customization operations were essential to meet participants' needs: the \textit{Hide}, the \textit{Shortcut}, the \textit{Move}, and the \textit{Reorder}. P6 also enjoyed customizing colors to create alternative color modes (\cref{fig:example} is an example of customization on a news website). P8, who valued the \textit{Shortcut} operation, mentioned his desire for a more advanced operation: \textit{``I would like the tool to redirect me to the web page I want... because what I did [throughout the study] was to add shortcuts''}.

Overall, different users have different needs. For instance, P6, an expert, mentioned only needing the CSS injector despite using the available operations when necessary (as shown in \cref{fig:example}). However, most participants, including experts, preferred to use the built-in operations.

People had benefits from using GitUI. P5, who assumed he would not become a user, mentioned the pleasure of accessing his \textit{``cleaner''} personalized web pages. P9, despite \textit{``not taking much pleasure in customizing''} (due to the effort), confessed that the result was \textit{``quite beneficial''} for her.

\subsubsection{Challenges}
Participants' most significant challenges in customizing for the self involved understanding what they could or needed to improve in the UI and what was possible to do with the customization tool.

\paragraph{Habituation and Ideation}
Deciding what to personalize is challenging. People get used to existing UIs (\textit{``I am already used [to UIs]... so I do not need that much of customizing''} (P7)); however, for most, this does not mean that they do not want to personalize but rather have \textbf{difficulties in imagining possible modifications}: \textit{``opening a web page and thinking about how it could be improved is not easy''} (P3); \textit{``it is difficult to escape from what we are used to''} (P8). For P8, having access to, for instance,  templates (similar to those offered by presentation programs) would make it easier to visualize possible improvements.


\paragraph{UIs Conceptualization.} Non-experts, particularly, did not \textbf{understand what they could adapt in a UI} or GitUI's potential: \textit{``as I did not understand the potential of the tool, I became a little unsure about how to use it''} (P3). For P3, customization was easier with \textit{``directions or goals''}, like with the ones of the received requests, and visualizing customization examples, like UIs with different background colors, would help her understand the tool's potential.

To support people's familiarization process with a customization tool, P4 suggests progressively providing different customization operations. For instance, as people would get used to UIs, they would gain access to more complex operations, allowing for more control over UIs: \textit{``it would be good if we could progress across levels of control... to stimulate us to perform different operations (...) then [after getting used to those operations] we would want more (...) basically, this would be a game where people progress in the way they personalize UIs''}.


\subsection{Requesting Customization Assistance}
The adoption of the request creation feature was low -- only used by P3. P3 assumed her motivation to make requests was to \textit{``experiment''} with the feature and could not imagine herself using it in the future: \textit{``it makes more sense for the elderly or people with disabilities''}. P3 assumed, however, that it was good to access her template done by others.

To justify the lack of requests, participants mentioned that they prefer to customize themselves. We found six factors that explain this attitude: self-efficacy, sense of control, selflessness, time and effort, articulateness, and self-perception.

First, most participants (experts and non-experts) mentioned that if they have a problem to solve, they will always try to solve it by themselves and only then request assistance, if necessary. Hypothetically, this can be a consequence of the high \textbf{self-efficacy} level of participants (\cref{table:participants}), as most mentioned having enough confidence in their skills to not ask for assistance: \textit{``I believe in my skills (...) if I can not do it, other people would not be able to do it either''} (P8).


Second, people enjoy being in \textbf{control}. We know that feeling in control \cite{marathe2011drives} represents a trigger to customize. Having control is making the customization decisions, including what and how to customize. Requesting customization assistance can make people lose part of their sense of control. P6 and P8 mentioned: \textit{``I like to do things myself''}.

Selflessness is the quality of thinking more about other people's needs than about your own. 
The influence of \textbf{selflessness} was evident with P5, who assumed he would not request assistance because he did not want to overload others. P5 only needs help with complicated operations that he can not perform. A request system would only be an option for him if he could make the requests directly to the website owners/developers.

Participants also mentioned the \textbf{time and effort} necessary to request customization assistance. P5 believes that he would only ask for assistance to perform an extensive redesign, which would be challenging as he would need to \textit{``invest one day to think about the design of the new version of the website''}. P8 mentioned that to request what he \textit{``really desires''} would make him lose as much time as if he customizes himself.

People must also describe their needs quickly and clearly (i.e., \textbf{articulateness}). P4 mentioned an expected difficulty  in describing her needs and revealed concerns that others would not understand her requests. The participant compared the received requests with the ones she could make, which seemed \textit{``more generic''}.

Finally, participants' \textbf{self-perception} regarding GitUI's users leads to conceptualizing two groups, those who answer requests and those who make them: \textit{``I belong to the group that does not ask for help''} (P2).


\subsection{Supporting Others to Customize}
People enjoyed the experience of supporting others to customize. There were no impacting challenges for participants, and each dealt with the different degrees of complexity of requests in their own way. For instance, experts injected CSS rules in a fictional request to hide YouTube thumbnail previews, while non-experts manually used the hide operation in all thumbnails. For more confusing requests, P9 confessed the need for a \textit{``visual clue''} (e.g., a screenshot with an illustration), and P9 doubts about implementing some solutions: \textit{``I thought `how to reply with the tools I have?'\,''}.

Most people assumed they would install GitUI only to assist others. They created strategies and reserved part of their day to reply to requests (e.g., \textit{``if the tool were available, I would like to receive a request every morning''} (P1)). Conversely, P5, with fewer replies, believes that users should learn how to customize themselves instead of asking for assistance.

\subsubsection{Motivation: The Daily Challenge and the Emotional Reward}\label{motivationforothers}
Two factors particularly motivated people: the challenge and the emotional reward (e.g., gratitude or connection with others).

Half of the participants mentioned the concept of challenge. They see a request as a challenge someone is making, and their motivation is to solve that challenge. P6 and P7 called it the \textbf{\textit{``daily challenge''}}. P4 mentioned that \textit{``the challenge itself was the motivation to reply to requests (...) it is challenging''}. P4 assumed not thinking about the requester and explained: \textit{``I saw each request as something to overcome... if it helps someone even better... but it is more like a game''}.

Another motivation was the \textbf{positive feeling of helping others}, for some, similar to the feeling of helping others in a physical context: \textit{``it is funny to use the tool and feels good to help others''} (P2); \textit{``the requests were simple things that I did quickly... and I felt good... even if it is just a first, basic, template so that later others can update''} (P1).

We also aimed to understand if accessing the requester's profile would influence if and how people reply to requests. For most, it had no influence. However, for P2, accessing the profile motivated her: \textit{``I like to know the information about whom I am helping (...) I imagine a face behind the request''}. The same goes for P9: \textit{``it makes the request more personal... it is not the details [profile] that matter (...) I even save templates with the requester's name... it generates a good feeling''}. Additionally, knowing the requester's profile can help prioritize requests (in a situation where multiple are available): \textit{``I would always prioritize requests with more impact... for example... helping blind people''} (P2).

\subsubsection{Leveraging User Engagement}
To support volunteers better and improve integration and motivation within the community, participants highlighted two main topics: \textbf{gamification} mechanisms and \textbf{feedback} on replies.

\paragraph{Gamification.}
Gamification mechanisms aim to drive user engagement in an activity \cite{inbookWood}. It includes components like points, badges, and leaderboards to represent and measure success, rewarding challenges, cooperation, or users' feedback. Participants mentioned that gamification components could be part of GitUI (e.g., \textit{``have you [the researcher] ever thought about including gamification?''} (P3); \textit{``would be great to have at least a star classification system''} (P8)). The first three participants proposed the gamification topic, so we included it in our interview script.

Interestingly, the suggested gamification mechanisms aim to leverage the motivation factors: the challenge and the emotional reward (\cref{motivationforothers}). First, people want to be \textbf{recognized for their successful replies}, for instance, by having a classification system. If they customize more and better than others, they want that information to be public and comparable. For instance, when the researcher told P6 that he needed one more reply to equalize the participant with more replies, he mentioned: \textit{``send me two requests then''}. P7, a \textit{fan} of the challenge of customizing for others, went further, mentioning the idea of \textit{``personalization teams''} competing with each other, \textit{``like teams of different universities''}.

Furthermore, participants suggested using \textbf{gamification components to allow requesters to demonstrate their gratitude and increase volunteers' emotional reward}. For instance, for P3, the most important is not to have a rating system but access her contributions: \textit{``I would like to see how many times I helped others (...) that would contribute to the positive feeling of contributing to something''}.

Overall, although all participants favored gamification mechanisms, they assumed it would not be a decisive factor in making them install a system like GitUI. For instance, P8, who assumed he would not become an GitUI user, mentioned: \textit{``[gamification] could not keep me motivated; I would always have other priorities''}. For P4, gamification components \textit{``can feed the ego (...) feed anyone's ego (...) but it is not the most important''}.

\paragraph{Feedback and Communication.}\textbf{Feedback} on replies was a crucial missing feature in GitUI. For P1, \textit{``even a simple message saying `your reply was accepted'\,'''} would motivate volunteers more and \textit{``generate gratitude''}. Also, for some, it would be good to communicate with the requester and ask for clarifications or feedback about the customization. For instance, due to the current lack of communication, P7 assumed that, while customizing for others, he customizes in his way (i.e., as he thinks he would like the result). However, the option to communicate with the requester lacks consensus. P3 believes that having an online chat could create \textit{``too high expectations on the requester''} (i.e., that replies would be perfect). For her, the best solution would be to have feedback on the replies, which the volunteer could use to improve the template. P3 confessed that it is \textit{``intriguing''} to be unsure whether she understood the received requests. For P4, who assumed requester profile access is nonessential, having access to feedback can help her get closer to the requester: \textit{``I would understand that there is a person on the other side''}.

\paragraph{Manage Volunteers' Confidence.} 
The notion that people are improving UIs when helping others can be a motivating factor. P9 mentioned: \textit{``after seeing the reply I made, I saw that it really improved the UI (...) it somehow motivated me... to have done something that turned out well''}.

Sending requests suited to people's abilities or predispositions to help can also be important to keep them motivated. Participants always tried to reply, even to more complex requests; however, P9 mentioned that when the request was easier, she felt more confident in her reply and motivated to reply to subsequent requests. She mentioned the desire to use GitUI \textit{``in a convenient way''} for her (i.e., reply to the more straightforward requests that make her feel good).

\subsubsection{Community and Social Requests}
One keyword mentioned by participants was \textit{community}. For instance, P4 mentioned a desire to be \textbf{part of a community}: \textit{``[the tool] can even form a community that works''}. To build this community, people mentioned the importance of being able to accept assistance requests and lock them (i.e., no one else could reply to that request). For P3, selecting and locking requests would provide a \textit{``feeling of greater control and freedom to volunteers''} by only committing when they are available and believing they can reply.

We also explored participants' reactions to the possibility of receiving requests directly from other users, like their friends. Overall, the idea was well accepted. For instance, P8, who assumed he would not be an active volunteer, is more open to personalizing for friends and family: \textit{``I would reply to all''}. P5, who also would not assist others, assumed the possibility to reply to more personal requests: \textit{``for strangers, I feel like there is no connection''}.

\subsubsection{Influence on Customizing for the Self}
Customizing for others influenced the customization for the self. For P7, the \textbf{received requests worked as suggestions} of templates he could make for himself, particularly removing unnecessary UI elements and adjusting the remaining. P3 has a similar opinion: by reading the requests, she could understand what she could adapt for herself, have a more critical attitude towards UIs, and better understand the \textit{``tool's potential''}. Also, replying to requests helped her explore features that she otherwise would not. P3 mentioned that the requests provided part of the guidance that she needed. For P2, requests helped her to reflect on how to improve UIs to herself. Requests also impacted the customization decisions of P5 and P8, however, by reminding them that they were part of a study where they should customize for themselves.

\subsection{User Experience and Usability}
In this section, we report participants' experience with GitUI, including the first contact during the think-aloud training session and the aspects reported in the final interview. The study's goal was not to evaluate the customization tool but to collect helpful feedback for future customization solutions and improve our prototype.

\paragraph{Learning Curve.}
During the training session, participants successfully learned how to use GitUI and showed confidence in using it independently at home. They were initially anxious (they did not know what to expect from a UI customization tool). However, their confidence grew as they started reading the names of the operations and relating them to the task. To execute the task, most participants started with the \textit{Hide} operation, which they assumed to be the easiest. Overall, once people got used to the workflow and the meaning of operations, they quickly finished the tasks.

\paragraph{Workflow.} The most frequent problem occurred when participants started the task: they did not know how to select an element to apply an operation. For most, the first step to customize was to try to select the UI elements of the web page (e.g., the text) and wait for a popup to show and provide any information about possible customization operations. With help, participants understood that they should first select the operation. Participants quickly adapted to this workflow but assumed that selecting the element before the operation would be more intuitive.

\paragraph{Experimentation.} People did not know precisely the values of the properties they were trying to edit (e.g., the color or font size), so they felt the need for experimentation. Ideally, they would select an element and then experiment with the operation and values to apply.

\paragraph{Agility and Fluidity.} Participants sought agility. As with the workflow, people are used to the standards imposed by other software, such as, spreadsheet editors. They wanted to be able to select multiple elements to apply the same operation; or the opposite: select an operation and then apply it to multiple elements. It should also be possible to keep an element selected and apply multiple operations or apply propagatable operations inside a container. 

A term mentioned by participants was fluidity. First, the menu should stay open between operations or have a \textit{pin} button to let users decide whether it should close. Second, as soon as they define the properties of an operation (e.g., select a color), the operation should be automatically activated -- without having to press any other button.

\paragraph{Feedback and Guidance.} All participants mentioned the desire for more guidance and feedback. Particularly, while executing an operation, the mouse cursor could be adapted to provide feedback about the ongoing operation, and step-by-step instructions could guide users during the process.

\section{Discussion}
Our work explores opportunities for future research in citizen-led personalization, aiming to empower common citizens to shape their digital experiences and move away from sole reliance on developers, designers, or system-driven personalization. We propose this control shifting through community-based approaches, democratizing the access to personalized UIs. Supported by a custom-designed tool, GitUI, we explored the context, motivation, and challenges of customizing for the self, asking for customization assistance, and helping others to customize. While we investigated these concepts using Web UIs, we anticipate that participants' behavior will be similar in other conventional devices or UI categories, although they present different technical challenges. In this section, we discuss our main findings.

\subsection{Customization Challenges: Guidance and Ideation}
Participants' adoption of GitUI to customize for themselves or request assistance was limited. While our tool does not allow for all the desired operations (e.g., workflow optimization), it was not a technical issue preventing customization: people have difficulties understanding how their UIs can be improved or even realizing how to take advantage of the features provided by the tool. To this end, the received requests helped them to realize what is possible to customize, which suggests benefits from making part of these requests publicly visible and discussed (i.e., open to the community).

We found that \textbf{people need help with the ideation process} (i.e., making design decisions). Participants proposed potential solutions to aid in this process, including visualizing various layout alternatives to draw inspiration from and experimenting with different layouts or properties to identify one that aligns with their expectations. We believe that \textbf{customization templates} (that can be further edited and adjusted to personal preferences) can be a solution to deliver these layout alternatives and support experimentation. Templates could be generic, semi-automatic, or created by the community. Generic templates would be similar to the templates available when creating a new presentation with a presentation program, which generally inspire people. These could be initially randomly generated but improved as the system increases its knowledge about users. Automated approaches could be used to provide more accurate templates. Community templates could follow the concept of Stylish \cite{stylish}, where people can share, access, and use templates created by others. Regardless of its origin, further customizing a template would be crucial -- ensuring the necessary transparency \cite{jameson2007adaptive} and control \cite{marathe2011drives} -- for instance, by allowing people to see a list of all applied operations in that template and select some to import for their own template.

To support people's ideation process, and similar to what was mentioned by \citet{mackay1991triggers}, other two approaches deserved to be further studied: (1) \textbf{creating situations to allow users to reflect on their UIs} (the received requests provide these situations); and (2) \textbf{bringing users into contact with each other} to share customization ideas or requests, which can increase the applicability and usefulness of individual customization decisions \citep{mackay1991triggers}.

\subsection{Supporting and Fostering Customization Requests}
Our participants did not need to create requests. Before further exploring other motives behind the lack of created requests we observed, future studies should focus on supporting the guidance and ideation processes. If people do not desire to customize (for instance, if they clearly do not see how they can improve their UIs, the benefits of doing it, or, at least, possible interaction problems needing, for instance, optimization), they logically will not ask for assistance. Therefore, future work should focus on \textbf{increasing the need for customized UIs} by exploring the previously mentioned solutions to the ideation problem and making people aware of customization benefits.

The ideation problem does not devalue the benefits of further supporting users in communicating their requests, facilitating and streamlining the process. GitUI offers a straightforward UI to create requests, and the articulateness of the request's description was pointed out as an issue. Existing work showed that even for designers, communicating UI changes is challenging, resulting in the need for a feedback loop to visualize changes or refine descriptions \citep{10.1145/3411764.3445573}.

Today, users typically communicate to UI creators the desired design or behavior features through cues (e.g., ``thumb up/down'' buttons) \citep{10.1080/07421222.2015.1029394}, which are insufficient to communicate customization intentions. Future research can work on top of more advanced communication methods, like the ones from UI behavior retargeting research \citep{10.1145/3472749.3474796, 10.1145/3411764.3445573}, and use visual references to support communication. For instance, CoCapture \citep{10.1145/3411764.3445573} enables users to create and describe dynamic UI behavior using mockups. The mockups contain changes that users want to propose to UI creators or questions about existing UIs. These communication mechanisms -- used by experts to build generic UIs -- can be further explored to facilitate communication between experts and non-experts requesters and volunteers.

Additionally, the recent progress of chatbots, like ChatGPT, also opens up space for \textbf{human-machine collaboration}, for instance, helping users describe their requests, bridging (e.g., ``translating'') the communication between experts and non-experts, or even supporting the ideation process. \citet{inproceedingsFischer} showed that ChatGPT can support design thinkers, for instance, formulating and solving design challenges. \citet{10.1145/3615335.3623035} also found that ChatGPT responds well to brainstorming and ideation prompts but not so well to design prompts, suggesting it can be explored to support the formulation of design challenges, ideation, and communication, and leaving its implementation to human volunteers.

\subsection{Motivating Volunteer Work}
We found that \textbf{users are motivated to assist others to customize by the good feeling of being helpful and by the challenge it represents}. Helping others increases happiness \cite{post2014altruism}, improves mood, and reduces stress \cite{midlarsky1991helping}. We can further raise their motivation and confidence by including gamification mechanisms, particularly to leverage these factors and let them know how many users they have helped \cite{10.1145/1124772.1124915}. Also, we could transform customization replies into a game (similar to approaches to label images \cite{10.1145/985692.985733}).

Our findings align with the motivating factors from other forum-based systems. For instance, Stack Overflow \footnote{https://stackoverflow.com} users are mainly motivated to contribute to the community by intrinsic factors \citep{9625742}, which include helping others, reciprocity, and making an impact \citep{penoyer2018impact}. Nevertheless, Stack Overflow users still enjoy the presence of gamified incentive mechanisms, for instance, when sharing their reputation on other platforms \citep{stackoverflowinsights}.

We believe that the factors motivating people to participate in the open source movement \cite{heron2013open} can also be used to motivate people to become customization volunteers. Factors like perfecting expertise \cite{lakhani2003hackers, von2003open} or enhancing reputation \cite{bezroukov1999open} can be important for experts, and factors like altruism \cite{benkler2006commons}, expectation of reciprocity \cite{morgan2000cathedral}, or fun and enjoyment \cite{bezroukov1999open}, could motivate volunteers in general. Our study confirms that it is important to allow individuals to \textbf{choose their level of participation} \cite{anthony2005explaining} and to self-select their contributions \cite{fuchs2008don} (i.e., the freedom to choose requests mentioned by P3).

Similarly, future research should focus on \textbf{adapting the complexity of requests to volunteers}, which can also increase their confidence. Most participants were open to creating requests, but only for more complex customization operations, requiring expert knowledge. However, we should not waste non-experts' availability to volunteer work. For complex assistance requests, customization can use crowdsourcing concepts \cite{liu2013accelerating}, where people work together to accomplish one specific task divided into microtasks with different complexities.

\subsection{Impact of Expertise and Personality Traits} We aimed to understand how expertise and personality affect the customization for the self and others. Our results showed \textbf{no variation in how people with different expertise and personality decide to customize, ask for assistance, or help others}. People shared characteristics like self-efficacy or the need for the ``daily challenge'' that resulted in similar reactions to the requests received. Participants' behavior is a positive finding, as previous studies \citep{Sundar2010} revealed that less tech-savvy people could demonstrate negative attitudes when asked to customize.

However, we found differences while customizing. Expertise impacted UI conceptualization (e.g., understanding a web page's structure) and the expected agility from GitUI. For instance, while non-experts expected a customization workflow and interface more consistent with mainstream software (e.g., word processors), experts desired more advanced interaction techniques (e.g., keyboard shortcuts), or while experts recurred to CSS rules to perform repetitive operations, non-experts found the process tedious. Future work, stimulating more customization requests and with a larger sample of participants, should clarify these differences to optimize personalization software use.

\subsection{Technical Learning for Customization}
Our study followed a qualitative approach. As we recruited a small number of people, we complemented previous work with in-depth insights into what people expect from a customization tool (collected from two weeks of usage). Future solutions should focus on \textbf{supporting the process of experimenting with different values and properties}, \textbf{guide users} during the customization process (with visual cues), and, above all, allow for \textbf{personalized customization workflows}. For instance, people should decide whether they want to apply the same operation to multiple elements or multiple operations to the same element.

We encountered technical challenges that are useful for future research. The Web is a complex ecosystem where developers have total freedom to define the design, workflow, and implementation languages. Therefore, the main challenge is dealing with all the constraints this diversity imposes. Also, web pages are frequently updated, resulting in customization templates with a short life span. Solutions to detect and correct templates that no longer work or inform users about them should be studied. Another vital aspect is scalability: future solutions should avoid duplicated requests or automatically detect solutions already available to new requests (e.g., with text mining approaches).

\subsection{Implications for UI (Community-Based) Personalization}
We envision a community-based approach to personalization that fosters collaboration between individuals with different skills and expertise. This study was a first step. We wanted to understand if people can collaborate towards producing personalized interfaces and how a system can support that process. Now that we know that they are available to collaborate, we can discuss the lessons learned and how to support this community. \cref{table:summaryresults} summarizes our findings that inform future community-based personalization approaches and customization research in general.

\begin{table*}[h!]
\centering
\caption{Lessons learned according to each research question.}
\label{table:summaryresults}
\Description{Summary of results for each research question.}
\begin{tabularx}{\textwidth}{l X}
\toprule

\multicolumn{2}{c}{\textbf{Customize for the self -- context, challenges, and motivations (RQ1)}}\\
\textbullet & People's primary motivation to customize is simplifying and optimizing web pages.\\
\textbullet & Problem identification and solution ideation are the biggest challenges to customization.\\
\midrule

\multicolumn{2}{c}{\textbf{Request customization assistance -- context, challenges, and motivations (RQ2)}}\\
\textbullet & Access to customization assistance does not increase people's willingness to customize. This suggests that besides reducing the required customization effort, which would be mitigated by requesting assistance, people must understand personalization benefits and existing UI problems.\\
\textbullet & Complex customization problems can provoke people to request customization assistance.\\
\textbullet & Self-efficacy, sense of control, selflessness, time and effort, articulateness, and self-perception represent challenges to requesting customization assistance.\\
\midrule

\multicolumn{2}{c}{\textbf{Provide customization assistance (RQ3)}}\\
\textbullet & People intrinsically enjoy customizing for others.\\
\textbullet & The positive feeling of helping others and the challenge it represents are the main motivators for volunteer work.\\
\textbullet & Volunteers need feedback on replies and to discuss customization details when necessary.\\
\textbullet & Typical forum-based gamification mechanisms can be adapted to the customization assistance context by creating a reputation built on the number of people a user has helped.\\

\bottomrule
\end{tabularx}
\end{table*}

\subsubsection{A Community-based Approach to Personalization}
The idea behind a customization community is to allow people who need a solution for a specific problem to interact with a wide community that shares similar interests and try to obtain a suitable solution \cite{paterno2013end}. Our findings indicate that community-based personalization is feasible, well-accepted, and could be further studied. Following the concept of GitHub \cite{10.1145/2145204.2145396} or crowdsourced Web site components \cite{nebeling2012crowdsourced}, people could securely and transparently collaborate to customize or create templates and requests.

A community-based approach to customization should include at least three components: customization for the self, customization assistance, and a customization repository. Customization for the self needs to support people's main needs of increasing efficiency and optimizing existing UIs. Other activities could be concentrated in a community hub repository, where people could collaborate to discuss templates or requests. These discussions could be public (others may share the same problem) and have different users work together to find a solution.

Assistance requests could be addressed in the repository, allowing people to search for requests describing similar problems and volunteers to collaborate to find a solution. Requests should be categorized so that people willing to help can follow requests on specific topics or websites. However, it is unclear whether requests should be ``locked'', as suggested by P3. By having requests available to multiple volunteers, requesters could select the best reply.

We found that the concept of customization templates works well. Following existing solutions (e.g., \cite{stylish}), users should be able to publish any personal template, making it available to users looking for a similar solution. People could redesign and experiment with existing templates and apply multiple (complementary) templates to the same UI. 

As in distributed version control systems, users could clone, redesign, and publish templates (as a different branch) and build a reputation that rewards the quality and quantity of the templates, assistance provided, or assistance requests. All the shared templates should be tested for malware and have their code/operations visible to potential users.

\subsection{Limitations}
The low number of created requests limited our findings, preventing us from studying aspects like peoples' capacity to describe desired UI changes or possible ways to support that process. Nevertheless, the concept was well received, and most participants showed interest in helping others and asking for assistance when necessary, justifying the need for future exploration of the community-based UI customization concept.

Furthermore, most participants, whether experts or non-experts, were inclined to use our tool outside the study scope but primarily to provide customization assistance. In this scenario, the challenge and positive feeling of helping others are already positive incentives. However, fully comprehending long-term user engagement requires a study longer than two weeks, during which the introduction of gamification mechanisms can prove crucial.

The inability to identify and articulate personalization needs challenges our community-based approach, which relies on individuals' intentions to customize, individually or through the community. Nevertheless, these challenges are not unique to our approach; they extend to user-driven personalization in general, underscoring the necessity for future research to support the identification of personalization opportunities. Also, performing a study with a longer duration can offer additional perspectives. For instance, most Stack Overflow users participate less than once per month, although most still feel part of the community \citep{stackoverflowinsights}.

GitUI was exclusively developed for Google Chrome, which is not the main browser of P2, P6, and P7. These participants assumed that this reduced their motivation to use the customization tool.

With our work, we do not pretend to devalue the work of developers or designers but instead call their attention to the importance of customization. Also, a request system can be beneficial in allowing website owners to understand possible interaction problems.

\section{Conclusion}
In this work, we explored how people customize for themselves and how they react to being able to request customization assistance or help others to customize. We went deep into this concept by performing an exploratory study with nine participants, where they independently used a customization tool, GitUI, for two weeks.

Our results suggest that participants are more open to customizing for others than for themselves. People need more guidance and help with the ideation process to customize for themselves. To customize for others,  people are motivated by the positive feeling of helping others and the challenge it represents. Future work should study mechanisms to help people customize for themselves or create requests (solving the challenges of ideation and guidance) and solutions to keep them further engaged in assisting others.

\begin{acks}
We thank the participants in our user study. This project was supported by Fundação para a Ciência e a Tecnologia through LASIGE Research Unit funding refs. UIDB/00408/2020 (https://doi.org/10.54499/UIDB/00408/2020), UIDP/00408/2020 (https://doi.org/10.54499/UIDP/00408/2020) and SFRH/BD/146847/2019, and Project 41, HfPT: Health from Portugal, funded by the Portuguese Plano de Recuperação e Resiliência.
\end{acks}

\bibliographystyle{ACM-Reference-Format}
\bibliography{sample-base}

\appendix

\newpage
\section{Results of the NEO-FFI-20}\label{ipipneo}
\begin{table}[h!]
\caption{Participants' big five personality traits (mean). Results from the NEO-FFI-20 questionnaire.}
\begin{tabular}{c|c|c|c|c|c}
\toprule
\rowcolor{white}
\textbf{ID} &\textbf{Neuroticism} & \textbf{Extraversion} & \textbf{Openness} & \textbf{Agreeableness}  & \textbf{Conscientiousness}\\ 
\midrule
\textbf{P1} & 2.75 & 2.75 & 2.5 & 3	& 2.75\\
\rowcolor{lightgray!50}
\textbf{P2} & 0 & 1 & 1 & 2 & 3\\
\textbf{P3} & 1.25 & 1.75 & 1.75 & 1.5 & 2.75\\
\rowcolor{lightgray!50}
\textbf{P4} & 2.75 & 2.75 & 2.25 & 1.75 & 3.25\\
\textbf{P5} & 2.75 & 1.75 & 1.75 & 2.50 & 3.25\\
\rowcolor{lightgray!50}
\textbf{P6} & 2.5 & 2.5 & 2.75 & 2 & 3.75\\
\textbf{P7} & 2.5 & 2 & 2.25 & 1.5 & 2.75\\
\rowcolor{lightgray!50}
\textbf{P8} & 2.5 & 2 & 1.5 & 1.5 & 3.25\\
\textbf{P9} & 2.25 & 1.25 & 1.75 & 1.25 & 3.5\\
\bottomrule
\end{tabular}
\end{table}

\end{document}